\documentclass[twocolumn,amsmath,amssymb,aps,lngbibliography,pra,floatfix]{revtex4-2}
\usepackage{graphicx}   
\usepackage{dcolumn}    
\usepackage{bm}         
\usepackage{hyperref}   

\usepackage{mathtools}
\usepackage{physics}
\usepackage[english]{babel}

\usepackage[T1]{fontenc}
\usepackage{siunitx}

\renewcommand{\vec}[1]{\mbox{\boldmath ${#1}$}}

\begin{document}    
    \title{Dynamics of Correlations and Entanglement Generation in Electron-Molecule Inelastic Scattering.}
    \author{Martin Mendez}
        \affiliation{Facultad de Matemática, Astronomía, Física y Computación, Universidad Nacional de Córdoba and Instituto de Física Enrique Gaviola, CONICET-UNC, Ciudad Universitaria,Córdoba X5000HUA, Argentina}
    \email{pont.federico@unc.edu.ar}
    \email{martinmendez@unc.edu.ar}
    \author{Federico M. Pont}
    \affiliation{Facultad de Matemática, Astronomía, Física y Computación, Universidad Nacional de Córdoba and Instituto de Física Enrique Gaviola, CONICET-UNC, Ciudad Universitaria,Córdoba X5000HUA, Argentina}
    
    \date{\today}                            
    
    \begin{abstract}

The dynamics and processes involved in particle-molecule scattering, including nuclear dynamics, are described and analyzed using various quantum information quantities throughout the different stages of the scattering. The main process studied and characterized with the information quantities is the interatomic coulombic electronic capture (ICEC), an inelastic process that can lead to dissociation of the target molecule. The analysis focuses on a one-dimensional transversely confined $\text{NeHe}$ molecule model used to simulate the scattering between an electron  $\text{e}^-$(particle) and a  $\text{NeHe}^+$ ion (molecule). The time-independent Schrödinger equation (TISE) is solved using the Finite Element Method (FEM) with a self-developed Julia package \hyperlink{https://github.com/mendzmartin/FEMTISE.jl}{FEMTISE} to compute potential energy curves (PECs) and the parameters of the interactions between particles. The time-dependent Schrödinger equation (TDSE) is solved using the Multi-configuration time-dependent Hartree (MCTDH) algorithm.
The time dependent electronic and nuclear probability densities are calculated for different electron incoming energies, evidencing elastic and inelastic processes that can be correlated to changes in von Neumann entropy, von Neumann mutual information and Shannon mutual information. The expectation value of the position of the particles, as well as their standard deviations, are analyzed along the whole dynamics and related to the entanglement during the collision and after the process is over, thus highlighting the dynamics of entanglement generation.
It is shown that the correlations generated in the collision are partially retained only when the inelastic process is active.
\end{abstract}

\keywords{TISE, TDSE, FEM, DVR, MCTDH, ICEC.}

    \maketitle
    
    \section{INTRODUCTION}
The study of particle-molecule scattering processes, including nuclear degrees of freedom, is fundamental for understanding chemical reactions, molecular dissociation, particle capture, as well as photochemistry~\cite{sisourat_interatomic_2010,otto_rotational_2008,arnold_control_2018,haxton_multiconfiguration_2011,albrecht_approximation_2023,palacios_theoretical_2015,pont_impact_2024,dey_quantum_2022,saalfrank_molecular_2020}, among other fields of application. The interaction between colliding particles and molecular ions, in the low kinetic energy range, extending up to several tens of electron volts (eVs), {provides insights into} the internal quantum mechanical behavior {and information transfer between the constituents, treated as subsystems}. The dynamical behavior of entanglement and correlation measures between these subsystems in charge migration~\cite{schurger_differential_2023}, photoexcitation by lasers~\cite{blavier_time_2022,saalfrank_molecular_2020}, and nuclear pathway manipulations~\cite{arnold_control_2018,dey_quantum_2022} is a topic of current interest, as it explores the fundamental physics underlying the transfer and generation of entanglement between subsystems in femtochemistry and attochemistry. Similar interests {also arise in} the capture and emission processes of charge carriers {within low-dimensional solid-state nanostructures}~\cite{buscemi_entanglement_2006,pont_electron-correlation_2016}.

There is a plethora of possible outcomes from a collision process, ranging from elastic scattering to dissociation mechanisms triggered by the excitation of the molecule or its constituent atoms~\cite{pedersen_electron_1999,alessi_state-selective_2012,barrachina_classical_2012,otto_rotational_2008,gokhberg_interatomic_2010,eckey_resonant_2018,grull_relevance_2022,grull_influence_2020,remme_resonantly_2023}. Inelastic scattering processes significantly contribute to generating correlations both among the different atoms within the molecule and with the scattered particle~\cite{tal_translational_2005}. A theoretically predicted inelastic process, referred to as interatomic Coulombic electronic capture (ICEC)~\cite{gokhberg_interatomic_2010,remme_resonantly_2023,jacob_interatomic_2019}, has garnered significant interest for its ability to induce molecular dissociation through the capture of an incoming electron. This process illustrates the impact of particle capture on molecular stability, particularly when nuclear dynamics are considered, and serves as a model for investigating entanglement dynamics in quantum scattering~\cite{tal_translational_2005,mack_dynamics_2002,arrais_entanglement_2016,o_osenda_dynamics_2008,buscemi_entanglement_2006}.

In this work, we investigate the dynamics of entanglement and correlations during the one-dimensional scattering of an electron ($\text{e}^-$) with a $\text{NeHe}^+$ ion. This one-dimensional system, derived from first principles, enables the examination of ICEC and related phenomena through a rigorous computational approach. Our study focuses on characterizing the various scattering stages using quantum information measures such as von Neumann entropy and mutual information, and Shannon mutual information~\cite{jaeger_quantum_2007}. Analyzing the electronic and nuclear probability densities over time at various electron impact energies reveals the presence of elastic and inelastic processes, with changes in quantum information quantities linked to these occurrences. Additionally, particle positions and standard deviations are connected to variations in entanglement and correlation throughout the collision process, elucidating their generation and determining whether they persist after the particle-molecule interaction is over.

The work is organized as follows, in Section~\ref{sec:model_description} the model for the confined NeHe$^+$ molecule is introduced, in Section~\ref{sec:methods} the selection of the different parameters of the model, the methods used for the computation of the potential energy curves and dynamical simulations, and the quantities used in the analysis of the results are described. In Section~\ref{sec:results} the description and discussion the results for the different quantities is presented to finally conclude in Section~\ref{sec:conclusion}.
    \section{MODEL DESCRIPTION}\label{sec:model_description}

\begin{figure}[!htbp]
    \centering
    \includegraphics[width=0.48\textwidth]{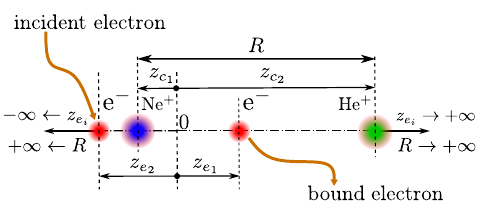}
    \caption{
    Schematic representation of the coordinates used in the electron-impact model, where {a projectile electron collides with the $\text{NeHe}^{+}$ ion}. The coordinate {$z_{e_2}$ represents the longitudinal position of the incoming projectile electron}, while {$z_{e_1}$ indicates the longitudinal position of the bound electron, and $R$ denotes the internuclear distance within the target}. The spatial arrangement {depicts the projectile electron on the left and the target ion on the right. The target consists of a Neon cation positioned to the left of the nuclear center of mass (CM) and a Helium cation situated to the right of the CM}.}
    \label{fig:eNeHe_molecule}
\end{figure}

In order to describe how {information measures can be used to track the dynamics of a scattering mechanism in a composite system}, a one-dimensional model for electron impact on NeHe$^{+}$ is used. The charged molecule is described by {two cations ($\text{Ne}^{+}$ and $\text{He}^{+}$) and an electron that binds them together}. The incident electron {collides with the molecule, resulting in elastic and inelastic scattering phenomena that may lead to various excitations or even dissociation of the molecule}. It is assumed that the system {is subjected to a transverse harmonic confinement potential, with the interactions between the cations and electrons derived in Sec.\ref{sec:eff1d}}. {A diagram of the system is shown in Figure \ref{fig:eNeHe_molecule}}.

The three-dimensional Hamiltonian operator for two electrons and two cations {subject to} lateral confinement can be written as,
\begin{equation} \label{eq:totham}
    \hat{H}_{\text{tot}} =\hat{T} +\hat{V}_{\text{conf}} +\hat{V}_{\text{int}},
\end{equation}
where $\hat{T}$ is the kinetic energy operator, $\hat{V}_{\text{conf}}$ is the confinement potential and $\hat{V}_{\text{int}}$ is the interaction potential. Explicitly, each term has the following form,
\begin{equation} \label{eq:hamterms}
    \hat{T} =\sum\nolimits _{i}\hat{T}_{i} ;\ \hat{V}_{\text{conf}} =\sum\nolimits _{i}\hat{V}_{i} ;\ \hat{V}_{\text{int}} =\sum\nolimits _{i,j\land i\neq j}\hat{V}_{ij}
\end{equation}

\noindent were $i,j=\{e_{1} ,e_{2} ,c_{1} ,c_{2}\}$ and $e$ refers to electron particle and $c$ to a cation particle. Each term in Eq.~\eqref{eq:hamterms} can be further expanded as
\begin{eqnarray}
    \hat{T}_{i} &=&\frac{\vec{p}^2_{i}}{2m_{i}} \\
    \hat{V}_{i} &=&\frac{m_{i}( \omega _{i})^{2}}{2}(\vec{\rho }_{i} \cdotp \vec{\rho }_{i}) \\
    \hat{V}_{ij} &=&\frac{q_{i} q_{j}\exp( -\alpha _{ij}| \vec{r}_{j} -\vec{r}_{i}| )}{| \vec{r}_{j} -\vec{r}_{i}| }
\end{eqnarray}
\noindent where $\vec{r}_{i}$ is the position vector and $\vec{\rho}_{i}$ is its polar projection. The expression for $\hat{V}_{i}$ represents a transversal harmonic confinement. The expression for $\hat{V}_{ij}$ represents generic electrostatic Yukawa (Coulomb for $\alpha_{ij}=0$) interactions between a pair of particles with charges $q_i$ and $q_j$, and specific parameters, defined by $\alpha_{ij}$, which are established as in Sec.~\ref{sec:parameters}.

The Hamiltonian~\eqref{eq:totham} can be rewritten in terms of nuclei center of mass (CM) and relative coordinates $\vec{r}_{\text{CM}}=(\sum_{i=c_1,c_2}m_i\vec{r}_{i})/m_{\text{tot}}$ and $\vec{r}_{c_{12}}=|\vec{r}_{c_1}- \vec{r}_{c_2}|$, respectively. The corresponding masses (total and reduced, in that order)  and canonical momenta are  $m_{\text{tot}}=(m_{c_1} + m_{c_2})$, $m_{\text{red}} = m_{c_{1}}m_{c_{2}}/m_{\text{tot}}$ and $\vec{p}_{\text{CM}}=(\vec{p}_{c_1}+ \vec{{p}}_{c_2})$, $\vec{p}_{c_{12}}= (m_{c_2}\vec{p}_{c_1} - m_{c_1}\vec{p}_{c_2})/m_{\text{tot}}$. Using this transformation we have,
\begin{equation}
\hat{H}_{\text{tot}} = \hat{T}_{\text{CM}} + \hat{H}_{3D}
\end{equation}

\noindent where
\begin{equation}\label{eq:ham}
\hat{H}_{\text{3D}}=\hat{T}_{c_{12}} + \hat{T}_{e_1} + \hat{T}_{e_2}+\hat{V}_{\text{conf}} +\hat{V}_{\text{int}}    
\end{equation}

\noindent is the working Hamiltonian. Note that the vectors $\vec{r}_{e_{i}} -\vec{r}_{c_{j}}$ can be rewritten as,

\begin{eqnarray} \label{eq:cmframe}
(\vec{r}_{e_{i}} -\vec{r}_{c_{j}}) &=& \vec{r}_{e_{i}} -\left[\frac{m_{\text{red}}}{m_{c_j}}(\vec{r}_{c_{1}} -\vec{r}_{c_{2}}) +\vec{r}_{\text{CM}}\right] \nonumber \\
&=&  \vec{r}_{e_{i}\text{CM}} -\frac{m_{\text{red}}}{m_{c_j}}\vec{r}_{c_{12}}
\end{eqnarray}

\noindent where $\vec{r}_{e_{i}\text{CM}}=\vec{r}_{e_i}-\vec{r}_{\text{CM}}$ is the coordinate of electron $i$ relative to the center of mass.

\subsection{Effective one-dimensional model}\label{sec:eff1d}
    
If the confinement potential is sufficiently strong (see Sec.~\ref{sec:parameters}) the total wave function can be approximated as
\begin{equation}
    \Psi (\vec{r}_{e_{1}} ,\vec{r}_{e_{2}} ,\vec{r}_{c_{12}}) = \psi (z_{e_{1}}, z_{e_{2}}, z_{c_{12}})\prod _{i=\{e_{1} ,e_{2} ,c_{12}\}}  \zeta^{(i)}_{0}( \vec{\rho}_i)
\end{equation}

\noindent where $\zeta^{(i)}_{0}(\vec{\rho}_i)$ are the eigenfunctions of
\begin{equation}\label{eq:hamconf}
    (\hat{T}_{\vec{\rho}_i} +\hat{V}_{i}) \zeta^{(i)}_{0}( \vec{\rho}_i) =\hbar \omega _{i} \zeta^{(i)}_{0}(\vec{\rho}_i)
\end{equation}
i.e., the ground state of a two-dimensional isotropic quantum harmonic oscillator, which represents the transversal part of the total wave function and $\psi(z_{e_{1}}, z_{e_{2}}, z_{c_{12}})$ is a one dimensional wave function (in each coordinate) representing a longitudinal part of the total wave function. The oscillator frequencies $\omega_i$ are set up considering a fixed confinement length size $l_c$ for all the particles.

The interaction energy between particles $k$ and $l$ can be expressed by the following integral,
\begin{equation}\label{eq:intpot}
    \left\langle \Psi | \hat{V}_{kl} | \Psi \right\rangle = \iiint \Vert \psi (z_{e_1},z_{e_2},z_{c_{12}})\Vert ^{2} V^{(kl)}_{\text{eff}}(| z_{kl}| )\prod _{i}dz_{i}
\end{equation}

\noindent where $V^{(kl)}_{\text{eff}}(| z| ) =\iiint  \vert\zeta^{(i)}_{0}(\vec{\rho}_{i})\vert^{2} V_{kl}(| \vec{r}| )\prod _{i}d\vec{\rho}_{i} $. As is shown in Eq.~\eqref{eq:intpot}, the interaction potential is expressed as the integral of the longitudinal wave function times an effective one dimensional potential energy $\hat{V}^{(kl)}_{\text{eff}}$~\cite{bednarek_effective_2003,pont_electron-correlation_2016}. A similar result is obtained for all the terms in the Hamiltonian~\eqref{eq:ham}, hence the following one dimensional effective Hamiltonian is obtained,

\begin{equation}\label{eq:hameff}
\hat{H}=\sum_i (\hat{T}_{z_i} + \hbar \omega_i) +\hat{V}_{\text{eff}},   
\end{equation}

\noindent where $\hat{V}_{\text{eff}}=\sum_{k\neq l} \hat{V}^{(kl)}_{\text{eff}}$. 
The effective interaction potentials $\hat{V}^{(kl)}_{\text{eff}}$ are analytically obtained for each pair of particles~\cite{wick_range_1938},

\begin{eqnarray}
    V_{\text{eff}}^{(kl)}( z) &=& \frac{\pi ^{\frac{3}{2}} q_{k} q_{l} }{2\left(\sqrt{2} l_{c}\right)}\text{exp}\left[\left(\frac{|z| }{\sqrt{2} l_{c}}\right)^{2} +\left(\frac{\sqrt{2} l_{c}}{2l_{\alpha_{kl}}}\right)^{2}\right] \nonumber \times \\ 
    & & \times \text{erfc}\left[\mathnormal{\frac{| z| }{\sqrt{2} l_{c}} +\frac{\sqrt{2} l_{c}}{2l_{\alpha_{kl} }}}\right], \label{eq:effpots}
\end{eqnarray}

\noindent where $q_k$ is the effective charge, $l_{\alpha_{kl}}=(\alpha_{kl})^{-1}$ is the Yukawa length and $l_{c}= \sqrt{\frac{\pi}{m_i \omega_i}} \quad \forall \quad  i=e_1,e_2,c_{12}$ is the confinement length. All parameter values are established in Sec.~\ref{sec:parameters}.

\section{METHODS}\label{sec:methods}

The simulation of the quantum dynamics of the collision process {can be divided, like any dynamical problem, into three steps: initial state computation, wave function propagation, and result analysis}. {Each of these steps is detailed in this section}. Moreover, {the interaction parameters must first be defined; therefore, this section begins by explaining the criteria used to select the parameter values for the effective one-dimensional model}.

\subsection{Parameters of the potentials}\label{sec:parameters}

The values of the parameters used in the model are tabulated in Table~\ref{tab:paramselection}.
\begin{table}[h]
    \caption{\label{tab:paramselection} Values of the parameters used in the simulations.}
    \begin{ruledtabular}
    \sisetup{table-number-alignment = center}
        \begin{tabular}{l l S}
            Symbol & Description & {Value(a.u.)} \\
            \hline
            $l_c$ & confinement length & 1.250 \\ 
            $l_\alpha$ & Yukawa length & 1.986 \\ 
            $q_{\text{He}}$ & effective charge of $\text{He}^+$ cation & 1.453 \\ 
            $q_{\text{Ne}}$ & effective charge of $\text{Ne}^+$ cation & 1.307 \\ 
            $\beta$ & cation interaction factor & 0.800 \\ 
            $m_{\text{Ne}}$ & Neon atom mass & 36785.339 \\
            $m_{\text{He}}$ & Helium atom mass & 7296.293 
        \end{tabular}
    \end{ruledtabular}
\end{table}

The studied system {is restricted to a quasi-one-dimensional region through harmonic confinement, as described in Eq.~\eqref{eq:hamconf}}. {All particles are confined within a region characterized by a length $l_c$, which determines distinct oscillator frequencies $\omega_i$ for each particle}. {The confinement length $l_c$ is calculated as the geometric mean of the covalent radius of Neon and the atomic radius of Helium}. {It is important to ensure that $l_c$ is sufficiently small so that the energies of the confinement's excited states can be neglected; this is verified after parameter selection and energy computation}.

{The effective electronic charge is set as $q_{e_i}=1 \text{ a.u.}$}. {The Yukawa length $l_\alpha$ and the effective charge $q_{\text{He}}$ are determined by aligning the ground state energy of the one-electron Hamiltonian $\hat{H}_{\text{He}}= \hat{T}_{e_1} + V^{(e_1 c_{\text{He}})}_{\text{eff}}$ with the first ionization energy of the Helium atom, $\epsilon_{\text{dis}}^{\text{He}}=-0.904 \text{ a.u.}$}. {This approach yields a curve $q_{\text{He}}(l_{\alpha})$ for the effective charges, as computed in Sec.\ref{sec:femtise}}~\cite{pont_impact_2024}. {A specific point on the curve is chosen such that the radial size of the charge density approximates that of Helium's bound state}. {The same method is applied to determine the effective charge for Neon, but in this case, the Yukawa length $l_\alpha$ is fixed to the value obtained for Helium}.

\subsection{PECs}

Even though Eq.~\eqref{eq:tdse} is solved { for the Hamiltonian of Eq.~\eqref{eq:hameff}} including full electron-nuclear dynamics, it is useful, for better understanding, to construct a Born-Oppenheimer (BO) approach and compute potential energy curves (PECs) for the target molecule $\text{NeHe}^{+}$. The PECs are the electronic state energies computed for each fixed distance between nuclei $z_{c_{12}}$. The molecule and coordinates of each particle are depicted in Fig.~\ref{fig:eNeHe_molecule}. From now on the name $R\equiv z_{c_{12}}$ {is used} for this internuclear distance, as it is the common use in BO literature. The PECs are the energies $\epsilon_n(R)$ computed as a function of the distance $R$ of
\begin{equation} \label{eq:TISE_1e1N}
\hat{H}_{\text{NeHe}^{+}}(R) \xi_n(z_{e_1};R) =  \epsilon_n(R) \xi_n(z_{e_1};R), 
\end{equation}

\noindent where $\hat{H}_{\text{NeHe}^{+}}(R)$ is the one electron Hamiltonian,
\begin{equation} \label{eq:hamiltonian_1e1N}
\hat{H}_{\text{NeHe}^{+}}(R) = \hat{T}_{e_1} + \hat{V}^{(e_1 c_{\text{He}})}_{\text{eff}}(R) + \hat{V}^{(e_1 c_{\text{Ne}})}_{\text{eff}}(R) +\hat{V}^{(c_{\text{Ne}} c_{\text{He}})}_{\text{eff}}(R).
\end{equation}

\noindent {where the parameters defined in Table~\ref{tab:paramselection} are used in the interactions}. The $R$ dependence in the effective terms for electron-nuclei interactions are explicitly included since they depend on the nuclei coordinates $(z_{\text{Ne}}$ and $z_{\text{He}})$, which can be expressed as (in the same way as in Eq.~\eqref{eq:cmframe}),
\begin{equation}
    z_{\text{Ne}} = -\frac{m_{\text{red}}}{m_{\text{Ne}}} R; z_{\text{He}} = \frac{m_{\text{red}}}{m_{\text{He}}} R,
\end{equation}

\noindent and using the masses from Table~\ref{tab:paramselection} we obtain,
\begin{equation*}
    z_{\text{Ne}} \equiv z_{c_{1}} = -0.165 R; z_{\text{He}} \equiv z_{c_{2}} = 0.834 R.
\end{equation*}

The PECs for the nuclear distance $R$ in $\text{NeHe}^{+}$ are shown in Fig.~\ref{fig:adiapot_and_eigenstates}. They were computed using the numerical approach described in Sec.~\ref{sec:femtise}. The ground state PEC shows a minimum which is located at the equilibrium distance $R_{\text{eq}}$. However, the obtained value is far from the reported value~\cite{seong_hene_1999} $\left( R_{\text{eq}}^{\text{exp}} =2.70 \text{ a.u.} \right)$, hence a multiplicative factor $\beta$ is included in the effective cation-cation interaction ($V^{(c_{\text{Ne}} c_{\text{He}})}_{\text{eff}}$) in order to correct this effect. The value of $\beta$ is selected such that the inter-nuclear coordinate $R_{\text{eq}}$ is as close to the experimental equilibrium distance as possible (this further explored in Sec.~\ref{sec:pecs}).

The PECs are also needed for the computation of the initial states used in the simulations (see Sec.~\ref{sec:initial_state}). Specifically, in the BO approach one has to solve the time-independent Schrödinger equation (TISE) for the nuclei coordinate $R$ with a Hamiltonian that includes the PEC as the only acting potential~\cite{gatti_applications_2017},

\begin{equation}\label{eq:radial_eq}
  \left(-\frac{\hbar^2}{2m_{red}} \frac{\partial^2}{\partial R^2}+ \epsilon_n(R)\right) \chi_n(R)=\tilde{E}_n \chi_n(R).
\end{equation}

\subsection{Numerical approach to solve the TISE}\label{sec:femtise}

FEMTISE~\cite{FEMTISEpackage} is a Julia~\cite{julia} self-developed package to resolve the TISE by finite element method (FEM). This is an implementation and extension over GRIDAP~\cite{gridap} package using high performance protocols and ARPACK library to efficiently compute the generalized eigenvalue problems. Strictly speaking FEMTISE finds the solutions of a weak formulation associated to original TISE which is a special case of Sturm-Liouville differential equation. In a nutshell, considering the FEM approach, which is a special case of Galerkin methods, we can numerically implement the weak problem as a generalized eigenvalue matrix problem~\cite{sun_finite_2016}. Then, using high performance algorithm, the package calculates a specific eigenpair, i.e. eigenstates and eigenenergies associated with a specific Hamiltonian operator. A remarkable consequence of the FEM is that, since this method is based on a variatonal formulation, the energies obtained by FEMTISE are upper bounds of the exact energies. The package is under active development and open access, it currently can solve one and two dimensional problems for arbitrary potentials. Main specific features of the package include: multi-thread and multi-tasks parallelization; simulation of pre-defined common potentials or simulation of two particles with different masses in one-dimension; also, the computation of eigenenergies as a function of an arbitrary potential parameter is easily done.

The following one-dimensional systems were computed with FEMTISE in this work: $\text{He}$ atom, $\text{Ne}$ atom and $\text{NeHe}^+$ ion to determine the interaction parameters selection and compute PECs as a function of the nuclei distance $R$. The simulations were configured using the settings described in Table~\ref{tab:FEMTSEsettingsimul}.

\begin{table}[h]
    \caption{\label{tab:FEMTSEsettingsimul} Settings for the simulations done using FEMTISE package.}
    \begin{ruledtabular}
        \begin{tabular}{llr}
            Symbol & Description & Value (a.u.)\\
            \hline
            $\Delta z_{e_1}$ & finite element size & 0.4 \\
            $(z_{e_1}^{\text{min}},z_{e_1}^{\text{max}})$ & electronic grid domain range & (-300,300) \\
            $\text{tol}$ & accuracy of eigenpairs output & $10^{-9}$ \\
            $\text{iter}$ & maximum number of iteration & 500 \\
            $\text{nev}$ & total number of eigenpairs & 500
        \end{tabular}
    \end{ruledtabular}
\end{table}

\subsection{Quantum dynamics of the collision}\label{sec:dynamics}

\subsubsection{Initial state}\label{sec:initial_state}

In ICEC, the incoming electron is captured by one moiety of the target 
system and an electron is emitted from another moiety, with a characteristic energy of the process. The initial state for the whole system has an
incoming state for the electron and a target initial state for the $\text{NeHe}^{+}$ molecule.
Since both electrons are identical, properly symmetrized electronic spatial wave functions, according to the spin projection of the electrons are selected,

\begin{eqnarray}\label{eq:initial_state}
\psi_{^{\text{sym}}_{\text{asym}}} &=& \left[\phi_{i}(z_{e1})\Phi_{\text{NeHe}^{+}}(z_{e_2},R) \right. \\ & & \left. \pm \phi_{i}(z_{e_2})\Phi_{\text{NeHe}^{+}}(z_{e_1},R) \right]\frac{1}{\sqrt{2}}. \nonumber
\end{eqnarray}

\noindent The incoming state $\phi_{i}(z)$ is a Gaussian shaped wave packet with incoming momentum $p_i$ (energy $\varepsilon_i$) and width $\Delta z_{e}$. $\Phi_{\text{NeHe}^{+}}(z_e,R)$ is a relaxed state obtained by imaginary time propagation~\cite{kosloff_direct_1986,manthe_wavepacket_1992} using an initial state $\xi_{0}(z;R_{\text{eq}})\chi_{0}(R)$, where $\chi_{n}(R)$ is defined in Eq.~\eqref{eq:radial_eq} and $\xi_{0}(z;R_{\text{eq}})$ is defined in Eq.\eqref{eq:TISE_1e1N}.

\subsubsection{Numerical approach to solve the TDSE}

The quantum dynamics is dictated by the time dependent Schrödinger equation (TDSE),

\begin{equation}\label{eq:tdse}
i\hbar \frac{\partial \psi}{\partial t} = \hat{H} \psi,
\end{equation}

\noindent where $\hat{H}$ is defined in Eq.~\eqref{eq:hameff}. The evolution of the system has an initial state described in Eq.~\eqref{eq:initial_state}. The actual evolution was performed using a Multiconfigurational Time dependent Hartree (MCTDH) algorithm. The algorithm assumes that our state can be described at all times during the evolution by an expansion in \emph{single particle functions} (SPFs) of the form,

\begin{eqnarray}
    \psi(z_{e_1},z_{e_2},R) = \sum_{j_1,j_2,j_3} A_{j_1 j_2 j_3}(t) \times \nonumber \\
    \varphi^{(e)}_{j_1}(z_{e_1},t)\varphi^{(e)}_{j_2}(z_{e_2},t)\varphi^{(N)}_{j_3}(R,t).
\end{eqnarray}

\noindent Note that each coordinate has its own \emph{time-dependent} optimized basis of SPFs $\left\{\varphi _{{j_i}}^{( \alpha)}( z_{\alpha} ,t)\right\}$. MCTDH theory and working equations for the coefficients $A_{j_1 j_2 j_3}$ and the SPFs can be found in the review~\cite{beck_multiconfiguration_2000}, several examples of application of the method to molecular quantum dynamics using the MCTDH-Heidelberg package~\cite{mctdh:MLpackage} can be found in the book~\cite{gatti_applications_2017}. 

The MCTDH algorithm has been extensively used to study quantum molecular dynamics~\cite{gatti_applications_2017}, quantum dynamics of bose condensates~\cite{fasshauer_multiconfigurational_2016,lin_mctdh-x_2020}, collision dynamics~\cite{otto_rotational_2008,pont_impact_2024}, electron dynamics in QDs~\cite{pont_electron-correlation_2016,pont_predicting_2020}, etc. The approaches used in the different applications have sometimes specific names as MCTDHF for fermions or MCTDHB for bosons. Here the regular (unsymmetrized) version of the algorithm as implemented in the MCTDH-Heideilberg package~\cite{mctdh:MLpackage} is used. However, the SPF basis for both electrons must be identical, since they are identical particles, and the coefficients are thus properly (anti-)symmetrized in the electronic indices. The electronic symmetrization of the initial state is implemented as described in Appendices~\ref{ap:initial_state} and~\ref{ap:propagation}. 

\subsubsection{CAPs}\label{sec:caps}

The collision problem studied here {differs from the molecular dynamics of closed systems without breakup reactions}. The main difference is the existence of electronic density far away from the target (the incoming electron), and a post-collision emitted electron density as well. In ICEC, {these two contributions are separated from the target molecule at all times except during the collision, which is a relatively short period}. Moreover, the collision {induces the dissociation of the $\text{NeHe}^{+}$ cation, which exhibits much slower dynamics}. Hence, {electrons are located far from the center of mass, requiring a very large grid to avoid unwanted reflections at the grid boundaries and accurately capture the dynamics}.

A partial solution to this issue is to include a complex absorbing potential (CAP) at the grid ends in each degree of freedom (DOF). A CAP is defined as,
\begin{equation}\label{eq:cap}
W^{(z)}_{\pm}=-i\eta (z \mp z_{\text{CAP}})^{2}\Theta(z \mp z_{\text{CAP}}),
\end{equation}

\noindent where $z_{\text{CAP}}$ is the starting point of the CAP, which extends up to the end of the grid. The effect of this potential is to absorb the density that enters this region. This density absorption can be used to compute the flux into that grid end~\cite{pont_electron-correlation_2016}. The absorbed density can quench the dynamics of the other DOFs, so one must be careful and locate the CAP far enough from the CM to prevent this effect.

\subsection{Correlation and entanglement measures used in the characterization and analysis of the dynamics}

\subsubsection{von Neumann and conditional von Neumann entropy}\label{sec:vnentropy}

The first and most direct analysis one can perform on the dynamics of the systems is to look at the electronic ($\rho_e$) and nuclear ($\rho_N$) densities,

\begin{eqnarray}
    \rho_e(z,t) &=&  \iint \left| \psi(z,z_{e_2},R,t)\right|^2  dz_{e_2}dR \label{eq:electronic_density} \\
    \rho_N(R,t) &=&  \iint \left| \psi(z_{e_1},z_{e_2},R,t)\right|^2  dz_{e_1}dz_{e_2} \label{eq:nuclear_density}
\end{eqnarray}

\noindent Note that in the case of the electronic density $\rho_e$, the integration can be over any one of the two electrons since they are identical.
Two-dimensional electronic density ($\rho_{e,e}$) and two-dimensional electron-nuclei density ($\rho_{e,N}$) such as,
\begin{eqnarray}
    \rho_{e,e}(z_{e_1},z_{e_2},t) &=& \int \left| \psi(z_{e_1},z_{e_2},R,t)\right|^2 dR \\
    \rho_{e,N}(z,R,t) &=& \int \left| \psi(z,z_{e_2},R,t)\right|^2 dz_{e_2}
\end{eqnarray}
\noindent are needed for the computation of the Shannon mutual information calculations of Section~\ref{sec:mutualinf}.

The natural orbitals are defined as the eigenstates of the electronic and nuclear reduced density matrices. These density matrices are defined as the partial traces,
\begin{eqnarray}\label{eq:reduced_densities}
    \hat{\rho}_e(t)&=&\text{Tr}_{R,z_{e_2}}(\hat{\rho}(t)), \nonumber \\
    \hat{\rho}_N(t)&=&\text{Tr}_{z_{e_1},z_{e_2}}(\hat{\rho}(t)),
\end{eqnarray}
\noindent where $\hat{\rho}(t)= \left|{\psi(t)}\right\rangle \left\langle{\psi(t)}\right|$ is the density matrix of the total system. The natural orbitals $\left| {\lambda^{(\alpha)}_{j}(t)} \right\rangle$ and their populations $\lambda^{(\alpha)}_{j}(t)$  are defined by the eigenvalue equation,
\begin{equation}\label{eq:nop}
    \hat{\rho}_{\alpha}(t) \left| {\lambda^{(\alpha)}_{j}(t)} \right\rangle = \lambda^{(\alpha)}_{j}(t) \left| {\lambda^{(\alpha)}_{j}(t)} \right\rangle
\end{equation}
\noindent where $\alpha=e,N$.
The natural orbitals are computed for each time step in the evolution and the norm of them is given by $\mathcal{N}(t)=\sum_{j=1}^{N} \lambda^{(\alpha)}_{j}(t)$, and might be less than one due to absorption by the CAPs (see Sec.~\ref{sec:caps}). They are also very important to test the convergence of the MCTDH method as, in the MCTDH approach, their number is the same as the number of SPFs. Analyzing the value of the least populated orbital population (LPOP) gives a bound on the convergence of the simulation. Hence, if the LPOP is rather high, one can augment the SPFs number and check whether the LPOP in the new simulation is as small as needed. In all simulations of this work the LPOP is less than $3 \times 10^{-6}$ for all times.

The von Neumann entanglement entropy (EE) of the reduced density matrices is computed using the natural orbital populations of Eq.~\eqref{eq:nop}, and has the following definition
\begin{equation}\label{eq:vNentropy}
    S_{\alpha}^{\text{vN}}(t) =-\sum_{j=1}^{N}\lambda^{(\alpha)}_{j}(t)\log_{2}\left(\lambda^{(\alpha)}_{j}(t)\right)
\end{equation}

The EE is a measure of entanglement for bipartite pure states~\cite{jaeger_quantum_2007,donald_uniqueness_2002,popescu_thermodynamics_1997}. Since the trace in Eq.~\eqref{eq:reduced_densities} is performed on two out three variables, the entanglement represented by Eq.~\eqref{eq:vNentropy} is between the two traced out and the other one. This means that $S_{N}^{\text{vN}}(t)$ corresponds to the nuclei-electrons entanglement and $S_{e}^{\text{vN}}(t)$ can be interpreted as a target-projectile entanglement in the collision dynamics. Note that, since these are von Neumann entropies of one part from a bipartition of a pure state, they are a measure of entanglement between the two parts~\cite{jaeger_quantum_2007}. There are two possible bipartitions in our system that lead to $S^{\text{vN}}_{e,e}\equiv S^{\text{vN}}_{N}$ and $S^{\text{vN}}_{e}\equiv S^{\text{vN}}_{e,N}$. Moreover, using the entanglement entropies the \emph{von Neumann conditional entropies} (CEs) of one and two particles are defined as~\cite{jaeger_quantum_2007},
\begin{eqnarray}
    S_{\alpha_1|\alpha_2}^{\text{vN}}(t)=S_{\alpha_1,\alpha_2}^{\text{vN}}(t)-S_{\alpha_2}^{\text{vN}}(t) \\
    S_{\alpha_{1},\alpha_{2}|\alpha_{3}}^{\text{vN}}(t)=S_{\alpha_{1},\alpha_{2},\alpha_{3}}^{\text{vN}}(t) -S_{\alpha_{3}}^{\text{vN}}(t)
\end{eqnarray}
\noindent Since the global quantum system remains in a pure state and its evolution is governed by unitary dynamics, it follows that $S_{\alpha_{1},\alpha_{2},\alpha_{3}}^{\text{vN}}(t)=0$. With these general definitions, the von Neumann CEs for the specific system treated here are,
\begin{eqnarray}\label{eq:condvNentropy}
    S_{e|N}^{\text{vN}}(t) &=&  -S_{e|e}^{\text{vN}}(t) = S_{e}^{\text{vN}}( t) -S_{N}^{\text{vN}}(t)\\
    S_{N|e}^{\text{vN}}( t) &=& 0 \\
    S_{e,e|N}^{\text{vN}}(t) &=& -S_{N}^{\text{vN}}(t) \label{eq:See.N}\\
    S_{e,N|e}^{\text{vN}}(t) &=& -S_{e}^{\text{vN}}(t) \label{eq:SeN.e}
\end{eqnarray}

Negative values of the CEs are a sufficient (not necessary) condition for the presence of entanglement~\cite{CerfNegativeEntropy}. The negative value obtained in Eq.~\eqref{eq:See.N} indicate the presence of entanglement between electrons (both taken as one part) and the nuclei coordinate, while Eq.~\eqref{eq:SeN.e} indicate the presence of entanglement between one electron and the nuclei-electron compound. The sign of Eq.~\eqref{eq:condvNentropy} is not defined a priori, however we will show that for the system and processes considered here, $S_{e}^{\text{vN}}( t) > S_{N}^{\text{vN}}(t)$ for all $t$ (see Figs. \ref{fig:symm_quantities_01_for_energy_04}, \ref{fig:symm_quantities_01_for_energy_08} and \ref{fig:symm_quantities_01_for_energy_12}), and thus the entanglement between electrons is confirmed ($S_{e|e}^{\text{vN}}(t)<0$).

\subsubsection{Shannon differential entropy}\label{sec:shentropy}

The Shannon differential entropies (SDE), continuous entropies or Shannon entropy in position space \cite{lin_shannon_2015} for a single electron and for the nuclei are defined as~\cite{rosenkrantz_brandeis_1989},
\begin{eqnarray}
    S_{e}^{\text{Sh}}(t) &=&-\int \rho_{e} \left(z,t\right)\log_{2}\left[ \rho_{e} \left(z ,t\right)\right] dz \label{eq:shaelec}\\
    S_{N}^{\text{Sh}}(t) &=&-\int \rho_{N} \left(R,t\right)\log_{2}\left[ \rho_{N} \left(R ,t\right)\right] dR. \label{eq:shaN}
\end{eqnarray}
\noindent The SDE for two electrons and for electron-nuclei are correspondingly defined as,
\begin{eqnarray}
    S_{e,e}^{\text{Sh}}( t) &=&-\iint \rho_{e,e}(z_{e_1},z_{e_2},t) \times \nonumber \\
     && \log_{2}\left[ \rho_{e,e}(z_{e_1},z_{e_2},t)\right] dz_{e_1}dz_{e_2} \\
    S_{e,N}^{\text{Sh}}(t) &=&-\iint \rho_{e,N}(z,R,t) \times \nonumber \\
     && \log_{2}\left[ \rho_{e,N}(z,R,t)\right] dzdR
\end{eqnarray}

The SDEs are the continuous analog of the discrete probability distributions used in the classical Shannon entropy~\cite{rosenkrantz_brandeis_1989,schurger_differential_2023} \footnote{The SDE is derived from a discrete probability of $N$ possible outcomes, assuming each probability corresponds to a point $x_i$ in a given interval $L$. Increasing the density of points, taking the limit $N\rightarrow \infty$, leads to a continuous probability density and the corresponding SDE. The complete expression for SDE includes two additive terms: a $\log(N)$ term and a $\log(L)$ (the measure) term~\cite{rosenkrantz_brandeis_1989}. The latter is formally needed for the argument of the logarithm to be adimensional. The SDE is, however, usually presented as in Eqs.~(\ref{eq:shaelec},~\ref{eq:shaN})}. The SDE decreases when the probability distribution (particle density, in this case) becomes more concentrated and increases when it spreads out. In general, a localized state has a lower SDE value than a delocalized one. 

\subsubsection{Mutual informations}\label{sec:mutualinf}

The electron-electron and the electron-nuclei von Neumann mutual informations (QMIs) can be defined using von Neumann EEs as~\cite{jaeger_quantum_2007},

\begin{eqnarray}
    I_{e:e}^{\text{vN}}(t) &=& 2S_{e}^{\text{vN}}(t) -S_{e,e}^{\text{vN}}(t) \nonumber\\
    &=& 2S_{e}^{\text{vN}}(t)-S_{N}^{\text{vN}}(t)  \\
    I_{e:N}^{\text{vN}}(t) &=& S_{e}^{\text{vN}}(t) +S_{N}^{\text{vN}}(t)-S_{e,N}^{\text{vN}}(t) \nonumber \\
    &=& S_{N}^{\text{vN}}(t) \label{eq:inf_mut_conditional}
\end{eqnarray}
\noindent The QMI measures the correlations (both quantum and classical) between two subsystems. The QMI is a non-negative quantity due to the subadditivity of von Neumann entropy \cite{Lieb_strong_subadditivity}. Note that the QMI exceeds the bound for the classical mutual information because quantum systems can be \emph{supercorrelated}~\cite{jaeger_quantum_2007}.

The electron-electron and the target-projectile Shannon mutual informations (SMIs) are similarly defined using the SDEs as~\cite{sagar_mutual_2005},

\begin{eqnarray}
    I_{e:e}^{\text{Sh}}(t) &=& 2S_{e}^{\text{Sh}}(t)-S_{e,e}^{\text{Sh}}(t) \\
    I_{e:N}^{\text{Sh}}(t) &=& S_{e}^{\text{Sh}}(t)+S_{N}^{\text{Sh}}(t)-S_{e,N}^{\text{Sh}}(t)
\end{eqnarray}
\noindent The SMI represents the degree of correlations between two continuous particle densities defined in \eqref{eq:electronic_density} and \eqref{eq:nuclear_density}. It has been previously studied in atomic systems ~\cite{sagar_mutual_2005}, where it is shown to quantify the interdependence between two particles, providing insight into how much information one conveys about the other. The SMI is always positive and becomes zero when the two-electron density or electron-nuclei density follows from a Hartree-type state. Thus, it can be interpreted as a measure of deviations from a Hartree-type reference state.
    \section{RESULTS AND DISCUSSIONS}\label{sec:results}

\begin{figure*}
    \includegraphics[width=1.0\textwidth]{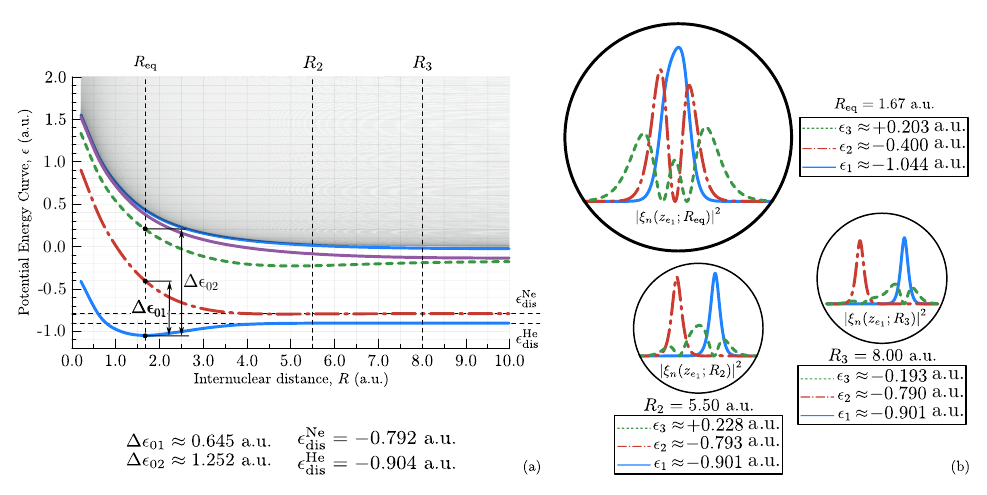}
    \caption{(a) Potential Energy Curves (PECs) of the one-dimensional $\text{NeHe}^{+}$ ion as a function of internuclear distance within the Born-Oppenheimer approximation. The first 500 eigenenergies are shown, with a grayscale gradient representing the continuum of energy levels. The horizontal dashed lines indicate the ionization energies of the Helium atom ($\epsilon_{\text{dis}}^{\text{He}}$) and of the Neon atom ($\epsilon_{\text{dis}}^{\text{Ne}}$) respectively. Furthermore, $\Delta \epsilon_{01}$ and $\Delta \epsilon_{02}$ are the energy differences at equilibrium distance between the ground state energy and the first and second excited state energies, respectively. (b) Electronic densities for the first three energy levels, obtained by solving the Time-Independent Schrödinger Equation (TISE) at selected internuclear distances. All calculations were performed using the FEMTISE package~\cite{FEMTISEpackage}.}
    \label{fig:adiapot_and_eigenstates}
\end{figure*}  

\subsection{PECs for $\text{NeHe}^+$}\label{sec:pecs}

The PECs for the $\text{NeHe}^{+}$ ion are computed by solving Eq.~\eqref{eq:TISE_1e1N} for different values of $R$ and are shown in Fig.~\ref{fig:adiapot_and_eigenstates}(a). Only the ground and first 2 excited states can bind vibrational nuclear states. Above the fifth curve, an accumulation of curves is developed that signals the pseudo-continuum onset (grayscale gradient). The dissociation limits ($R\rightarrow \infty$) of the ground (first excited) curve corresponds to the Helium (Neon) first ionization energy, and the corresponding electronic density is shown in Fig.~\ref{fig:adiapot_and_eigenstates}(b) for $R=8.00 \text{ a.u.}$. The states show that according to our setup in Fig.~\ref{fig:eNeHe_molecule}, the Helium is on the right and the Neon is on the left. The equilibrium distance for the ground state $R_{\text{eq}} \approx 1.67 \text{ a.u.}$ is below the experimental values reported for the molecule~\cite{seong_hene_1999}. The Yukawa interactions proposed here aim to model a one-dimensional system with tunable interactions and to show the effect of electron-nuclear dynamics on correlations and entanglement and, within its limitations, the model correctly reproduces some characteristic of the $\text{NeHe}^{+}$ ion, but could not reproduce the $R_{\text{eq}}$ more accurately. 

The density corresponding to the second excited state near the equilibrium distance of this state is shown in Fig.~\ref{fig:adiapot_and_eigenstates}(b) for $R_2=5.50 \text{ a.u.}$. It shows how the electronic density is increased between the nuclei and provides binding. For longer distances, $R_3=8.00 \text{ a.u.}$, this state develops into an excited state for the Helium atom as is apparent from the inset.

The energy differences $\Delta \epsilon_{01}$ and $\Delta \epsilon_{02}$, between the ground state energy at $R_{\text{eq}}$ and the first and second excited state energies, are useful to determine the energy at which one expects the ICEC channel to be open. This is true in a fixed nuclei approach, however this activation energies are modified by the inclusion of nuclei dynamics as shown in Ref.~\cite{pont_impact_2024}. Moreover, new paths that can be quite different appear by inclusion of the nuclei dynamics and will be described in the next section.

\subsection{Scattering details}\label{sec:scattering_details}

This section depicts how to study the time dependent electronic and nuclear densities and how to spot different characteristics and scattering channels from the time evolution. The dynamics published in Ref.~\cite{pont_impact_2024}, using different interaction potentials, {were} focused on the main differences between fixed nuclei and full dynamics. The abbreviated analysis presented here is useful in the following sections where it is contrasted to the one performed using the information obtained from the EEs and SDEs.

\begin{figure*}
    \centering
    \includegraphics[width=1.0\textwidth]{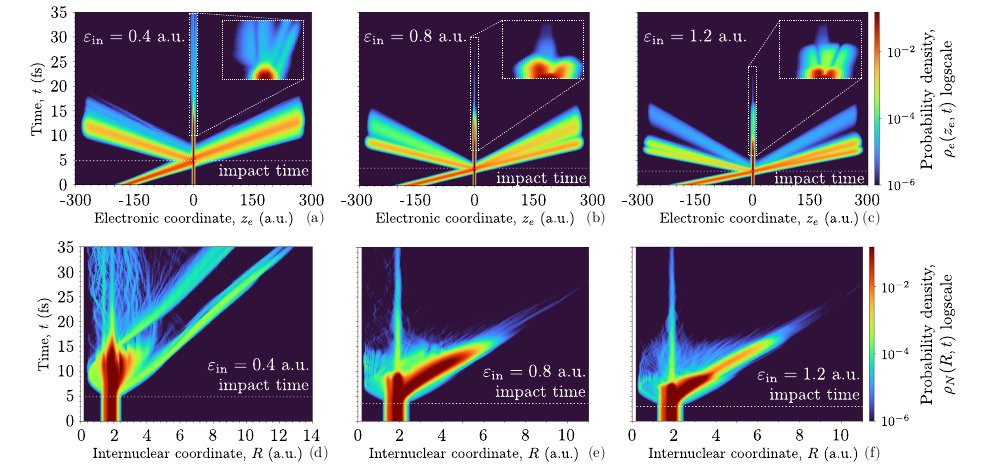}
    \caption{Electronic and nuclear probability densities at various incoming electron energies. The insets illustrate magnified views of the electronic probability density over specified spatial and temporal ranges: (a) electronic coordinate range from $-12$ to $12$ a.u. and evolution time from $10$ to $35$ fs, (b) electronic coordinate range from $-8$ to $8$ a.u. and evolution time from $7$ to $30$ fs, (c) electronic coordinate range from $-11$ to $11$ a.u. and evolution time from $6$ to $24$ fs. The horizontal dashed line marks the collision zone.}
    \label{fig:symm_cationic_density_energy_04}
\end{figure*}

The dynamics of the electronic scattering against $\text{NeHe}^{+}$ depends, mostly, on the energy distribution of the incoming wave packet. The effect of three different incoming energies is discussed: below the first excited state ($\varepsilon_{\text{in}} = 0.4\,\mbox{a.u} < \Delta \epsilon_{01}$), above the first excited state ($\Delta \epsilon_{02}> \varepsilon_{\text{in}}=0.4\,\mbox{a.u.} > \Delta \epsilon_{01} $) and near the second excited state vertical threshold ($\varepsilon_{\text{in}} = 1.2\,\mbox{a.u.} \approx \Delta \epsilon_{02}$). The energy dispersion of the projectile wave packet ($\Delta \varepsilon=0.06 \, \mbox{a.u.}$) is the same in the three cases. The time-dependent electronic and nuclear densities are shown in Fig.~\ref{fig:symm_cationic_density_energy_04}. {In all cases, the initial wave function was symmetric under electron exchange, corresponding to a singlet spin state.}

For the lower energy, the electronic density, Fig.~\ref{fig:symm_cationic_density_energy_04}(a), shows an incoming Gaussian shaped density from the left and impacting the $\text{NeHe}^{+}$ molecule (at the indicated impact time) and resulting in a reflected and transmitted densities. Both of these densities show no large changes in energy, which can be identified from the slope of the peak values of the densities. Only the characteristic spreading of the Gaussian wave packet in both cases is observed. Hence, only an elastic scattering channel seems to be open. However, the nuclear densities (Fig.~\ref{fig:symm_cationic_density_energy_04}(d)) give a deeper insight and make apparent a \emph{vibrational} excitation of the nuclei and, moreover, a dissociation channel of the nuclei. The vibrational excitation shows that there is an inelastic \emph{vibrational} channel, restricted to the ground state $\epsilon_{0}(R)$. The absorption of density by the CAPs at the grid edge makes apparent that the emitted electronic density is in the same channel to most of the vibrational excitations, since the nuclei density is strongly reduced at the same time that the electronic density is absorbed at the grid edge (around $15\text{ fs}$). There is, however, density that survives this absorption: A small vibrational density localized near the equilibrium distance and two dissociation branches (the two lines with similar slopes). The electronic density that is connected to this two contributions is shown by the inset in Fig.~\ref{fig:symm_cationic_density_energy_04}(a). The densities show that some density is bound to the molecule and the rest is evolving with the dissociation channel.  

The setup with energy $\varepsilon_{\text{in}} = 0.8 \, \mbox{a.u.}$ (see Figs.~\ref{fig:symm_cationic_density_energy_04}(b) and~\ref{fig:symm_cationic_density_energy_04}(e)) includes clear evidence of inelastic electronic emission. The process is detected by noting that after the collision we observe two different peaks for the electronic density with different slopes. One has the same energy as the incoming wave packet while the other is a slower electron emission with a slope compatible to the ICEC process to the first excited state. The nuclear density shows two main channels: an elastic density, that stays localized near $R_{\text{eq}}$, which is absorbed when the elastic electron reaches the grid edge (around $10$ fs) and a dissociation channel, that is absorbed by the time the ICEC electron reaches the grid edge (around $15$ fs). This is the clear indication that the dissociation is mainly due to ICEC. Specifically, the incoming electron is captured by the molecule in the dissociative first excited state and the excess energy is taken by the bound electron which is, in this case, emitted with a lower energy than the incoming one. The capture in the first excited state is made evident by the inset of Fig.~\ref{fig:symm_cationic_density_energy_04}(b), noting that the structure is that of the electronic state densities depicted in Fig.~\ref{fig:adiapot_and_eigenstates}(b) at $R_{\text{eq}}$.

The higher energy case, $\varepsilon_{\text{in}} = 1.2 \, \text{a.u.}$ (see Figs.~\ref{fig:symm_cationic_density_energy_04}(b) and \ref{fig:symm_cationic_density_energy_04}(e)), involves three different density \emph{branches}: the elastic channel (EC) with the smaller slope, the first excited state ICEC channel (ICEC$_1$) with a middle slope and the second excited ICEC channel (ICEC$_2$) with the higher slope. The EC and ICEC$_1$ channels are the most intense ones, and hence the two most probable processes to happen. The ICEC$_2$ is much more less intense, and hence less probable. EC and ICEC$_1$ have a higher probability in the forward direction than in the backward direction which contrasts with the rather equally probable forward and backward emission of ICEC$_2$. One important difference between ICEC$_1$ and ICEC$_2$ is the symmetry of the electronic density (see Fig.~\ref{fig:adiapot_and_eigenstates}(b)) which is asymmetric in channel 1 ($|\xi_2|^2$) and nearly symmetric in channel 2 ($|\xi_3|^2$). Moreover, the incident electron comes from the left and the capture into ICEC$_1$ eases the emission to the right of the bound electron. On the other hand, for ICEC$_2$ the effect is compensated by the fact that the second excited state density at $R_{eq}$ is surrounding the density for the ground state, giving an equally probable forward and backward emission.

\subsection{Dynamics of quantum information 
measures}\label{sec:quantinf_results}
The entanglement and correlation between the different bipartions of a physical system can be quantified using the von Neumann EEs and the SDEs, as well as mutual informations, as defined in Sections~\ref{sec:vnentropy},~\ref{sec:shentropy} and~\ref{sec:mutualinf}. Here we describe how {the different measures evolve in time}  during the collision process.

\subsubsection{Von Neumann entropy}\label{sec:vonneumann_results}
\begin{figure}[!htbp]
    \centering
    \includegraphics[width=0.48\textwidth]{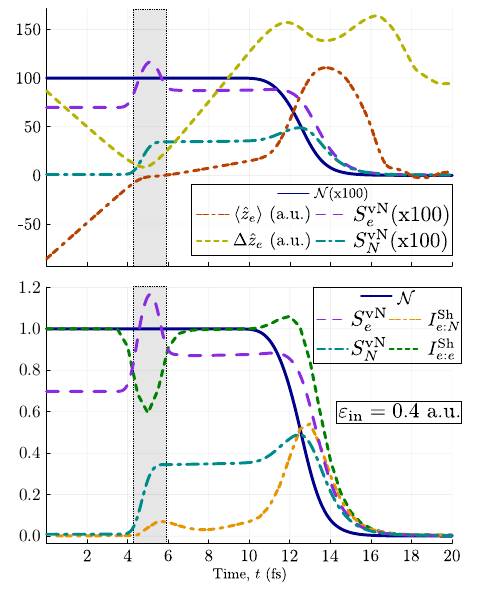}
    \caption{{ The top panel shows}, the expectation value of the electron position ($\left\langle 
\hat{z}_{e} \right\rangle$), the electronic position dispersion ($\Delta \hat{z}_{e}$), the target-projectile EE ($S_{e}^{\text{vN}}$), the nuclei-electron EE ($S_{N}^{\text{vN}}$) and the norm ($\mathcal{N}$). The bottom panel displays $S_{e}^{\text{vN}}$, $S_{N}^{\text{vN}}$, the electron-electron SMI ($I_{e:e}^{\text{Sh}}$), the target-projectile SMI ($I_{e:N}^{\text{Sh}}$) and the norm. { Both panels correspond to an incoming electron energy of $0.4 \text{ a.u.}$.} The shaded region indicates where the collision occurs. {Note that $\left\langle\hat{z}_e\right\rangle$ and $\Delta\hat{z}_e$ are computed using the normalized densities, while the entropies are computed using the densities without normalization.}}
    \label{fig:symm_quantities_01_for_energy_04}
\end{figure}

The results for the case of low incoming electron energy ($\varepsilon_{\text{in}}=0.4\, \text{a.u.}$) are discussed first. Figure~\ref{fig:symm_quantities_01_for_energy_04} shows the von Neumann EEs for the target-projectile, $S^{\text{vN}}_{e}$, and for the nuclei-electrons, $S^{\text{vN}}_{N}$, along with the norm,  electronic dispersion and electronic position.

The electronic dispersion $\Delta \hat{z}_e$, is rather large at the initial time up to values very close to the "collision time" {where} it shows a minimum. This is because the electronic state is properly \emph{antisymmetrized}, hence the probability density is double peaked, and the dispersion reflects this two peaked distribution. The minimum at the collision is not only due this two peaked distribution coming together, but also to a \emph{squeezing} of the incoming wave packet due to the strong repulsive interaction with the bound electron.  As presented here, the "collision time" is a rather loose concept, since it depends, as we will see on Sec.~\ref{sec:collision_time}, on the quantity one uses to define it.

Noticeable, the target-projectile EE ($S^{\text{vN}}_{e}$) develops a peak near this collision time \footnote{Actually, this can be another characteristic time that can be pinpointed as \emph{the collision time}}. More interesting is that, previous to the collision, $S^{\text{vN}}_{e}$ has a value corresponding to a maximally mixed electronic reduced density (because it includes the entanglement between the bound and incoming electron), and after peaking at the collision time ($\approx 5 \text{ fs}$) it stabilizes again to a higher value than the initial one (this may not be the case in two particle scattering, see Ref.~\cite{hahn_nonentangling_2012}). In other words, the initial entanglement measure is highly increased in the collision, but not all of it is retained. Thus a natural question that arises is how much of this entanglement is retained in different scenarios. We will discuss this in Sec.~\ref{sec:entanglement_generation}.

The collision effect is also visible in the nuclei-electron EE, since it {rises} from a nearly zero value up to a plateau. However, there is no peak in the entanglement during the collision. The entanglement increase clearly shows that during the collision the nuclei and electrons strongly correlate and that this entanglement is sustained in time.

The natural orbital populations $\lambda^{(\alpha)}_j$ (used to compute the von Neumann EEs, as shown in Eq.~\eqref{eq:vNentropy}) indicate that many electronic populations climb up to rather high values at the collision, and then decrease. This points to vibrations being temporarily excited during the collision. Also one can connect the wave packet squeezing at collision time to this raise and decrease of the populations. This is shown in the Appendix~\ref{ap:nop}.

After all collision effects are over, the emitted electron is absorbed by  CAP at the grid edge. This absorption leads to a {loss} in norm, as seen in Fig.~\ref{fig:symm_quantities_01_for_energy_04} from $10 \text{ fs}$ on. This has an important effect on the entropies and mean values, since the absorbed probability density of the particle (and its corresponding terms in the wave function expansion) are no longer part of the system, and one must be careful in the interpretation of the results. 

\begin{figure}[!htbp]
    \centering
    \includegraphics[width=0.48\textwidth]{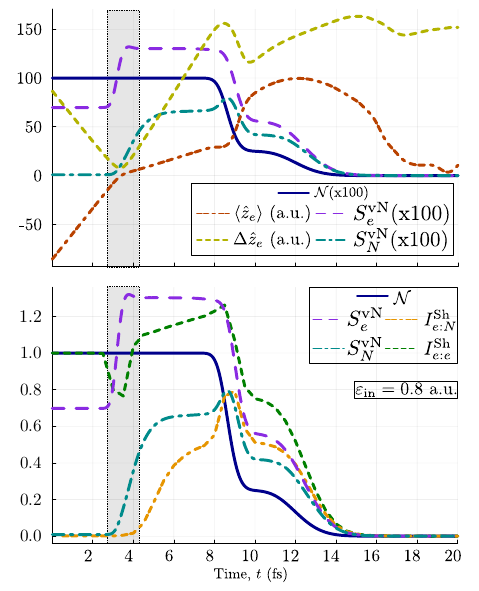}
    \caption{Same as in Fig.~\ref{fig:symm_quantities_01_for_energy_04}, for an incoming electron energy of $0.8 \text{ a.u.}$.
}
    \label{fig:symm_quantities_01_for_energy_08}
\end{figure}

\begin{figure}[!htbp]
    \centering
    \includegraphics[width=0.48\textwidth]{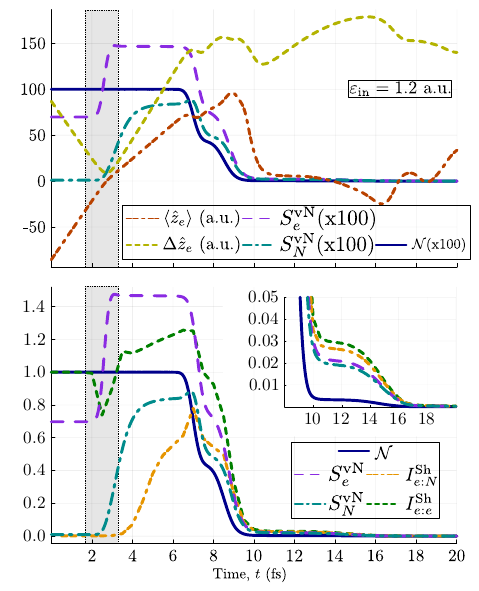}
    \caption{Same as in Fig.~\ref{fig:symm_quantities_01_for_energy_04}, for an incoming electron energy of $1.2 \text{ a.u.}$.
    Additionally, the inset provides an amplified view of the quantities for evolution times greater than $8\,\text{fs}$, where only $\text{ICEC}_{2}$ channel is left, this made apparent by a plateau in the plotted curve.
    }
    \label{fig:symm_quantities_01_for_energy_12}
\end{figure}

The changes in the entropies, due to the absorption  can be used to detect the different channels present for a given energy. Figs.~\ref{fig:symm_quantities_01_for_energy_04},~\ref{fig:symm_quantities_01_for_energy_08} and~\ref{fig:symm_quantities_01_for_energy_12} show the time dependent entropies for the three cases shown in Fig.~\ref{fig:symm_cationic_density_energy_04}. For $\varepsilon_{\text{in}}=0.4 \text{ a.u.}$, since the absorption is rather smooth and all of the entropy goes to zero, only the elastic channel is present.

In the case of $\varepsilon_{\text{in}}=0.8 \text{ a.u.}$, one can see that the absorption starts at $8 \text{ fs}$, and then has a small plateau at $10 \text{ fs}$, after which it decays again until it vanishes. This behavior matches exactly the absorption of the two present channels: the elastic and ICEC to the first excited state. The decay between 8 and 10 fs corresponds to the absorption of the elastic channel, after which, a small period of time with no absorption follows. This small plateau, which lives until the slower ICEC electron reaches the CAP, gives the entropy contribution of ICEC. The entropy decays again after the plateau, at a different pace than in the elastic channel.
This analysis shows that the combination of entropy and CAPs in this setup enables the detection of different channels. This is particularly evident for $\varepsilon_{\text{in}}=1.2 \text{ a.u.}$, where the system exhibits elastic scattering as well as ICEC$_1$ and ICEC$_2$ emitted electrons. These three absorptions are identifiable in the entropy measures, revealing distinct differences in absorption times and plateaus for each channel.

\subsubsection{Mutual informations}\label{sec:mutualinf_results}

\begin{figure*}
    \centering
    \includegraphics[width=1.0\textwidth]{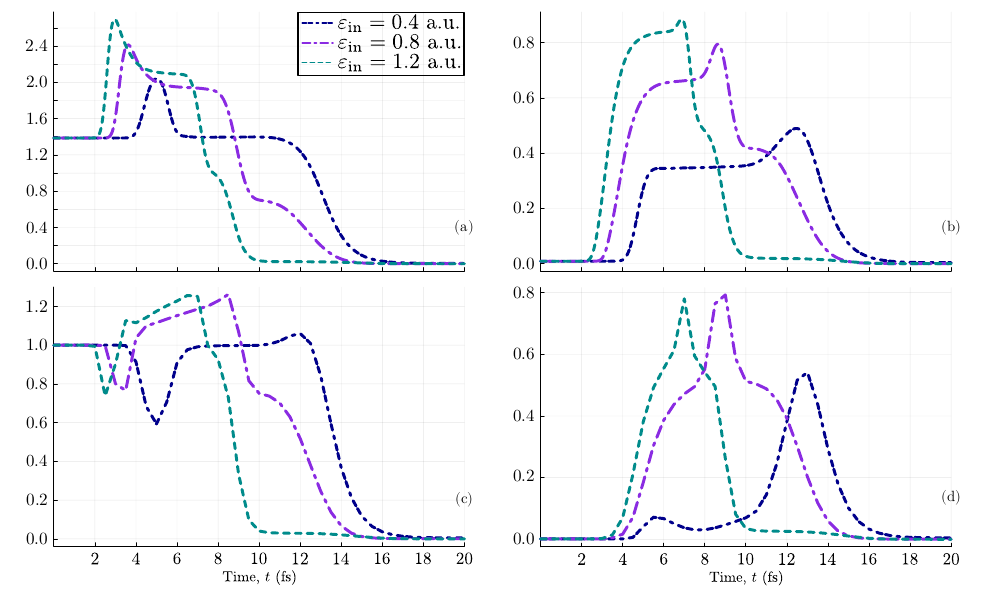}
    \caption{Time evolution of mutual informations for different electron incoming energies: (a) electron-electron QMI, $I_{e:e}^{\text{vN}}$, (b) electron-nuclei QMI, $I_{e:N}^{\text{vN}}$, (c) electron-electron SMI, $I_{e:e}^{\text{Sh}}$, and (d) electron-nuclei SMI, $I_{e:N}^{\text{Sh}}$.}
    \label{fig:symm_mutual_info_040812}
\end{figure*}

The mutual informations defined in Section~\ref{sec:mutualinf}, can be computed for the electron-electron or the electron-nuclei pair. Fig.~\ref{fig:symm_mutual_info_040812} shows the behavior of the SMI as compared to the QMI.

At the time of the collision, the electron-electron SMI ($I_{e:e}^{\text{Sh}}$) (see Fig.~\ref{fig:symm_mutual_info_040812}(c)) reaches a local minimum, indicating a reduction in electronic density correlations between the electrons. This implies both, that less information about one electron can be inferred by measuring the other and the departure of the state from a Hartree-type (product) one. In contrast, during the collision, both the target-projectile von Neumann EE ($S_{e}^{\text{vN}}$) (see Figs.~\ref{fig:symm_quantities_01_for_energy_04},~\ref{fig:symm_quantities_01_for_energy_08} and~\ref{fig:symm_quantities_01_for_energy_12}) and the electron-electron QMI ($I_{e:e}^{\text{vN}}$) (see Fig.~\ref{fig:symm_mutual_info_040812}(a)) show local maxima. The increase in $S_{e}^{\text{vN}}$ reflects an enhancement in target-projectile \emph{entanglement}, as discussed in Sec.~\ref{sec:mutualinf}. Similarly, the rise in $I_{e:e}^{\text{vN}}$ suggests that \emph{all correlations}, including enhanced correlations and non-locality between electrons~\cite{jaeger_quantum_2007}, intensify. Another noticeable difference between these quantities is that they refer to correlations in different subsystems. The entropy $S_{e}^{\text{vN}}$ describes entanglement between the target and projectile, while $I_{e:e}^{\text{vN}}$ refers to \emph{all correlations} between electrons. For the low energy case there are no new correlations between electrons, as the value of $I^{\text{vN}}_{e:e}$ is the same before and after the collision. Entanglement between electrons and nuclei is present, but not increased, since the nuclei are vibrationally excited in this case and the EE $S^{\text{vN}}_{N}$ is temporarily raised (note it is numerically identical to $I^{\text{vN}}_{e:e}$). For higher energies, correlations between electrons are build up due to the inelastic excitation of the bound electron, and also nuclei-electron correlations grow even more as they include vibrations and excitations.

After the collision time, when a certain percentage of the elastic channel density is absorbed by the CAP, both the electron-nuclei SMI ($I_{e:N}^{\text{Sh}}$) and the electron-nuclei QMI ($I_{e:N}^{\text{vN}}$) exhibit global maxima. This is because the density absorbed at the CAP corresponds to the elastic channel, which does not contribute to the mutual information (the mutual information is zero before collision). Hence, the mutual information of the remaining density is higher because it does not have the uncorrelated density contributions (the norm is decreasing, see Figs.~\ref{fig:symm_quantities_01_for_energy_04},~\ref{fig:symm_quantities_01_for_energy_08} and~\ref{fig:symm_quantities_01_for_energy_12}). The collision enhances electron-nucleus correlation . In addition, it is observed that $I_{e:N}^{\text{vN}}$, for high incoming electron energies, exhibits peaks that are greater than or equal in magnitude to those of $I_{e:N}^{\text{Sh}}$. This occurs because the von Neumann mutual information captures not only classical correlations but also quantum correlations. SMI on the other hand relates the correlations from the probability distribution density, which can be very small for localized states. At higher energies, more ICEC channels become available, which involve additional correlations. This information is seen to be most easily detectable through the QMI, as the SMI does not shows such a marked increase in these cases.

The nuclei-electron EE ($S_{N}^{\text{vN}}$), turns out to be equivalent to $I_{e:N}^{\text{vN}}$ in the present case (see Eq.~\eqref{eq:inf_mut_conditional}). However, $I_{e:N}^{\text{vN}}$ reflects all correlations between an electron and the nuclei, and not only entanglement between them. Also, the knowledge of one electron state is not a significant factor in quantifying nuclei-electrons correlations, because they are identical particles, and many of the equivalencies on the different quantities come from this fact.

\subsection{Impact time estimation and information measures}\label{sec:collision_time}

The collision event can be spotted using different quantities. For example, the electronic EE ($S^{\text{vN}}_{e}$) and the electron-electron QMI($I^{\text{vN}}_{e:N}$) both show maximum values within the collision region, while electronic dispersion ($\Delta \hat{z}_{e}$) and electron-electron SMI ($I_{e:e}^{\text{Sh}}$) show minimum values. There is need to properly define a collision time ($t_{\text{col}}$) in order to compare the entanglement measures at that particular time with the entanglement before ($t_{\text{be}}=t_{\text{col}} - \Delta t$) and after ($t_{\text{af}}=t_{\text{col}} + \Delta t$) the collision. The collision time can be estimated analyzing those quantum information measures in the collision region ($\mathcal{I}_{\text{col}}=[t_{\text{be}},t_{\text{af}}]$), that is,
\begin{eqnarray}
    t_{\text{col}}^{S} \text{ } : &S_{e}^{\text{vN}}\left(t_{\text{col}}^{S}\right) \ge S_{e}^{\text{vN}}(t) & \forall t \in \mathcal{I}_{\text{col}}, \\
    t_{\text{col}}^{\Delta} \text{ } : &\Delta \hat{z}_{e_{i}}\left(t_{\text{col}}^{\Delta }\right) \le \Delta \hat{z}_{e_{i}}(t) & \forall t \in \mathcal{I}_{\text{col}}, \\
    t_{\text{col}}^{I} \text{ } : &I_{e:e}^{\text{Sh}}\left(t_{\text{col}}^{I}\right) \le I_{e:e}^{\text{Sh}}(t) & \forall t \in \mathcal{I}_{\text{col}}.
\end{eqnarray}

\begin{figure}[!htbp]
        \centering
        \includegraphics[width=0.48\textwidth]{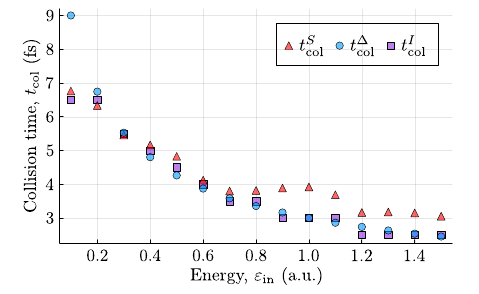}
        \caption{Estimated impact time as a function of incoming electron energies, determined by extreme values of von Neumann electronic EE ($t_{\text{col}}^{S}$), electron-electron SMI ($t_{\text{col}}^{I}$) and electronic dispersion ($t_{\text{col}}^{\Delta}$).}
        \label{fig:estimated_impact_time}
\end{figure}

The results are presented in Figure~\ref{fig:estimated_impact_time}. For low energies, below $0.2$ a.u., the collision time estimation by electronic dispersion is significantly higher than the time estimations by electronic SMI or by von Neumann electronic EE. This result occurs because the incoming electron energy is insufficient to produce a significant transmitted electronic density. As a result, most of the electronic density is reflected, leading to high dispersion in the collision region. Consequently, the collision is not clearly distinguishable, making estimation based on electronic dispersion less accurate.

At intermediate energies, from $0.2$ a.u. to $0.8$ a.u., all the estimated collision times are similar.

Finally, at high energies, above $0.8$ a.u., the estimation of collision time by the von Neumann electronic EE is slightly higher than the other time estimations. The essence of this different estimation is that the electronic SMI and the electronic dispersion measure correlations of the electronic system without taking electronic entanglement into account, while it is taken into account by the von Neumann electronic EE.

The von Neumann electronic EE is then best suitable as an estimator to give a proper "collision time" in this type of quantum collision dynamics using wave packets, since it includes quantum entanglement which is one of the characteristic features of the collision.

\subsection{Entanglement generation}\label{sec:entanglement_generation}

Once the collision time was estimated as described in \ref{sec:collision_time}, one is able to compute how much relative electron-nucleus entanglement is present as,

\begin{eqnarray}
    \delta_{S}(t)=\frac{S_{e}^{\text{vN}}(t)}{S_{e}^{\text{vN}}(t_{\text{col}})}.
\end{eqnarray}

It is  interesting to compare the entanglement before and after the collision with that at the collision time $t_{col}$, since during the collision, correlations increase significantly according to Figure~\ref{fig:symm_mutual_info_040812}. 

\begin{figure}[!htbp]
        \centering
        \includegraphics[width=0.48\textwidth]{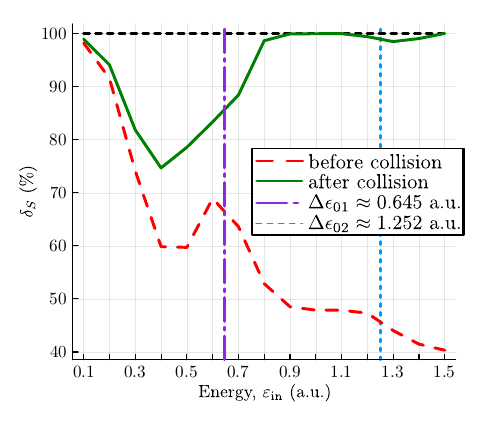}
        \caption{The relative von Neumann entropies ($\delta_S$) as a function of incoming electron energies. The plot shows the relative entropy values evaluated at times before ($t_{\text{be}}$) and after ($t_{\text{af}}$) the collision. Vertical dashed lines represent the energy differences at equilibrium distance between the ground state energy and the first excited state ($\Delta \epsilon_{01}$) and second excited state ($\Delta \epsilon_{02}$).}
        \label{fig:relative_entropy_value}
\end{figure}

There is an energy range (see Figure~\ref{fig:relative_entropy_value}), from $0.2$ a.u. to $0.7$ a.u., where from the total electron-nuclei entanglement generated at the collision time, about $20\%$ is lost after the collision. At higher energies, most of the entanglement is conserved after the collision. The reason is that there are many available channels at higher energies, and correlations to this states can be created easily,  keeping it inside the system. For lower energies, the number of accessible channels is  reduced, the transition matrix elements may also be smaller, and thus a part of the entanglement generated is lost because the system goes back to lower energy levels.

The relative value of entropy before the collision gives a hint on how much entanglement (as measured by $S_{e}^{\text{vN}}$) is generated. It also shows a region, from $0.5$ a.u. to $0.7$ a.u., where the generated entanglement is reduced from the trend. For $\varepsilon_{\text{in}}=0.5$ a.u., from the total amount of generated entanglement, about $60\%$ is initial entanglement, while for $\varepsilon_{\text{in}}=0.6$ a.u. this rises to $70\%$. The reason is that the energies are approaching the first excited state energy threshold $\Delta \epsilon_{01}$. A similar effect is seen for energies from  
$1.0$ a.u. to $1.2$ a.u., again matching to the onset of the second excited state threshold $\Delta \epsilon_{02}$. The result is that near the threshold values, less entanglement is generated during the collision.

    \section{CONCLUSION}\label{sec:conclusion}

The work analyzes entanglement and correlation measures of an inelastic electron capture and emission process (ICEC) by a dimer molecule, in which the nuclei movement is fully included. Long-range interactions (Coulomb potential) for electron-electron repulsion and  short-range interactions (Yukawa potential) for electron-nucleus attraction and nucleus-nucleus repulsion are used. The Yukawa potential simulates electron screening effects, and the electrons were treated as identical particles. Moreover, based on previous works~\cite{pont_electron-correlation_2016}, an effective one-dimensional potential is constructed to account for harmonic confinement. This confinement can also be used to model realistic boundary conditions in trapped ion systems or quantum dots, and it is also relevant for exploring phenomena such as quantum phase transitions and collective excitations in confined systems. 

The results from the probability densities give an understanding about collision time and zone, elastic and inelastic scattering, the symmetry of the electronic states and dissociation mechanisms. 
A novel approach to analyze the collision dynamics using quantum information theory measures, such as von Neumann EE, SMI and QMI, was introduced. The comparison with previously used quantities such as the differential Shannon entropy of the diagonal density matrix is also presented ~\cite{blavier_time_2022,schurger_information_2023}. These quantities enable the identification of the number and type of scattering channels (elastic and inelastic) during propagation, as well as the quantification of correlations between subsystems. Additionally, the collision time is estimated using three specific quantum information metrics.
The amount of entanglement preserved after the collision and the entanglement generated by the scattering process, as measured by the von Neumann EE, is discussed and seen to be highly connected to inelastic processes. The energy range where the inelastic processes are active shows that most of the entanglement is kept after collision. Electron-electron QMI has also shown to be a useful quantity, because it can detect the presence of inelastic scattering by revealing correlations (both quantum and classical) between the electrons that persist after the inelastic collision process is over. In addition to the main topics studied, a self-developed software package~\cite{FEMTISEpackage} (implemented in Julia) to compute potential energy curves (PECs) was developed. 

As a future line of work,  it would be of interest to implement a quantum discord measure~\cite{ollivier_quantum_2001} for these systems. For example, it is seen here that total correlations are increased or kept constant after the collision (see $I^{vN}_{e:e}$ in Fig. \ref{fig:symm_mutual_info_040812}) and also that electrons are entangled during the whole process $S^{vN}_{e|e}(t)<0$ for all $t$'s. The discord would enable to distinguish quantum correlations (entanglement and others) from the classical ones. Another interesting point about the ICEC process is the actual time it takes for the process to happen, since this would be an important tool to estimate whether ICEC can be measured in an experiment or not. An analysis of this particular point could be done using the present dynamical description, by setting up different initial conditions corresponding to experimental setups.
    \begin{acknowledgments}
 We gratefully acknowledges partial financial support of CONICET (PIP-KE311220210100787CO), SECYT-UNC (Res. 233/2020) and ANPCYT-FONCyT, PICT-2018 Nº 3431. M.M. acknowledges financial support by CONICET under a doctoral fellowship. This work used computational resources from CCAD – Universidad Nacional de Córdoba (\url{https://ccad.unc.edu.ar/}), particularly Mulatona and Serafin Clusters, which are part of SNCAD – MinCyT, República Argentina.
\end{acknowledgments}
    \section{Author contributions}

M.M. designed and developed the FEMTISE software and run all the calculations on MCTDH. M.M. and F.M.P. equally contributed to scientific conceptualization, formal analysis and manuscript writing.
    \appendix
\section{Building the initial state in MCTDH-Heidelberg package}\label{ap:initial_state}

The initial state for the quantum evolution is the properley symmetrized product state of the ground state of $\text{NeHe}^+$ ion model with one active electron times a gaussian for the projectile electron. The Hamiltonian for the ground state is given by,

\begin{equation} \label{eq:hamrelaxation}
    \hat{H}_{e_{1},R} = \hat{T}_{e_1} + \hat{T}_{N} + \hat{V}^{(e_1 c_{\text{He}})}_{\text{eff}} + \hat{V}^{(e_1 c_{\text{Ne}})}_{\text{eff}} +\hat{V}^{(c_{\text{Ne}} c_{\text{He}})}_{\text{eff}}
\end{equation}

\begin{table}[h]
    \caption{\label{tab:MCTDHrelax} Setting for $\text{NeHe}^+$ ion relaxation using MCTDH-Heidelberg package.}
    \begin{ruledtabular}
        \begin{tabular}{llr}
            Symbol & Description & Value [au]\\
            \hline
            $N_{z_{e}}$ & number of points for DVR grid& 1501 \\
            $N_{R}$ & number of points for DVR grid & 501 \\
            $(z_{e}^{\text{min}},z_{e}^{\text{max}})$ & electronic grid domain range & (-300, 300) \\
            $(R^{\text{min}},R^{\text{max}})$ & nuclear grid domain range & (0, 20) \\
            $\text{SPF}_{z{e}}$ & number of electronic SPFs & 20 \\
            $\text{SPF}_{R}$ & number of nuclear SPFs & 18 \\
            $\text{tol}_{\text{CMF}}$ & tolerance of CMF integrator & $10^{-3}$ \\
            $\text{tol}_{\text{RK8/spf}}$ & tolerance of RK8 integrator & $10^{-9}$ \\
            $\text{tol}_{\text{rrDAV/A}}$ & tolerance of rrDAV integrator & $10^{-8}$ \\
            $\text{tol}_{\text{eps inv}}$ & tolerance of eps inv integrator & $10^{-10}$ \\
            $t_{\text{final}}$ & final relaxation time & $82.684$~\footnote{It is equivalent to $2.0$ fs.} \\
            $t_{\text{out}}$ & step relaxation time & $0.413$~\footnote{It is equivalent to $0.01$ fs.} \\
        \end{tabular}
    \end{ruledtabular}
\end{table}

The \emph{relaxation} to the ground state of the Hamiltonian~\eqref{eq:hamrelaxation} is performed using imaginary time propagation as implemented in the MCTDH Heildeberg package~\cite{mctdh:MLpackage}. This relaxation is performed from specific ansatz for each DOF: for $z_{e_1}$ corresponds to the ground state of the Hamiltonian in Eq.~\eqref{eq:hamiltonian_1e1N}, except here the internuclear coordinate is fixed at the equilibrium value of the ground-state PEC $(R_{\text{eq}})$, for $R$ corresponds to the ground state of the Hamiltonian given by $\hat{H}_{R} = \hat{T}_{R} + \epsilon_{0}(R)$, where $\epsilon_{0}(R)$ refers to the PEC of the ground state (see Fig.~\ref{fig:adiapot_and_eigenstates}(a)). The simulations were performed using the configurations shown in Table~\ref{tab:MCTDHrelax}.

Once the ground state of the Hamiltonian~\eqref{eq:hamrelaxation} is obtained, the product with the Gaussian of the incoming electron must be included. Actually, due to the algorithm implementation, the electron incoming electron DOF, say $z_{e_2}$, is already included as a product to a Gaussian with no interaction to $R$ or $z_{e_1}$. 

In general terms,  in the MCTDH-algorithm, the wave function is written  as a Hartree product~\cite{mctdh:MLpackage},
\begin{eqnarray}\label{eq:hartree_expansion}
    \Psi ( q_{1} ,q_{2} ,\dotsc ,q_{n} ,t) =\sum _{j_{1} =1}^{n_{1}} \dotsc \sum _{j_{f} =1}^{n_{f}} A_{j_{1} \dotsc j_{f}}( t) \times \nonumber \\
    \times \prod _{k=1}^{f} \varphi _{j_{k}}^{( k)}( q_{k} ,t)
\end{eqnarray}

\noindent where $\left\{\varphi _{j_{k}}^{( k)}( q_{k} ,t)\right\}_{j_{k} =1}^{n_{k}}$ represents the single particle function (SPF) time-dependent basis for the degree of freedom (DOF) $q_k$, and $ A_{j_{1} \dotsc j_{f}}( t)$ are the time-dependent coefficients that control the phase selection of each (SPF).

In our case, we obtained the relaxation of the $\text{NeHe}^{+}$ ion, thus the initial wave functions at this stage of the simulation are given by:
\begin{eqnarray}
    \Psi _{\text{relax}}^{\text{ini}} &=& \sum_{j_{1},j_{2},j_{3}} A_{j_{1} j_{2} j_{3}}^{\text{relax}} \phi_{j_1}^{(e_1)}(z_{e_1}) \phi_{j_2}^{(N)}(R) \phi _{j_{3}}^{(e_{2})}(z_{e_2}) \nonumber \\
    &=& \sum_{j=1}^n A_{jj1}^{\text{relax}} \phi_{j}^{(e_1)}(z_{e_1}) \phi_{j}^{(N)}(R)
\end{eqnarray}
\noindent were we consider only a single trivial term in the projectile's SPF basis (with $n_3=1$ and $\phi_{1}^{(e_2)}(z_{e_2})=1$). The $A_{jj1}^{\text{relax}}$ coefficients are the only non-zero coefficients obtained from the relaxation, and we note that only the diagonal terms ($j_{1}=j_{2}=j$) persist.

After relaxation step, we define a Gaussian wave packet for the projectile, while maintaining the same Hartree product configuration for the target,
\begin{equation}
    \Psi_{\text{punch}}^{\text{ini}} = \sum_{j} A_{jj1}^{\text{relax}} \phi_{j}^{(e_1)}(z_{e_1}) \phi_{j}^{(N)}(R) \phi_{1}^{\left(\text{gauss}\right)}(z_{e_2}).
\end{equation}

We then need to build a symmetric state for the electronic coordinates and, by setting \emph{symorb=1,3} MCTDH keyword, a global SPF basis is constructed for the electronic system from the individual SPF basis associated to each electron as follows,
\begin{equation}
    \begin{pmatrix}
    \left\{\varphi _{j_{1}}^{( e_{1})}( z_{e_1})\right\}_{j_1=1}^{n_{1}} & \left\{\varphi _{j_{2}}^{( e_{2})}( z_{e_2})\right\}_{j_2=1}^{n_{2}} & \left\{\varphi _{j}^{(e)}(z_e)\right\}_{j=1}^{n}\\
    \phi _{1}^{( e_{1})}( z_{e_1}) & \phi _{1}^{( e_{2})}( z_{e_2}) & \phi _{1}^{( e)}(z_e) =\phi _{1}^{( e_{1})}\\
    \phi _{2}^{( e_{1})}( z_{e_1}) & \phi _{2}^{( e_{2})}( z_{e_2}) & \phi _{2}^{( e)}(z_e) =\phi _{1}^{( e_{2})}\\
    \phi _{3}^{( e_{1})}( z_{e_1}) & \phi _{3}^{( e_{2})}( z_{e_2}) & \phi _{3}^{( e)}(z_e) =\phi _{2}^{( e_{1})}\\
    \phi _{4}^{( e_{1})}( z_{e_1}) & \phi _{4}^{( e_{2})}( z_{e_2}) & \phi _{4}^{( e)}(z_e) =\phi _{2}^{( e_{2})}\\
    \phi _{5}^{( e_{1})}( z_{e_1}) & \phi _{5}^{( e_{2})}( z_{e_2}) & \phi _{5}^{( e)}(z_e) =\phi _{3}^{( e_{1})}\\
    \phi _{6}^{( e_{1})}( z_{e_1}) & \phi _{6}^{( e_{2})}( z_{e_2}) & \phi _{6}^{( e)}(z_e) =\phi _{3}^{( e_{2})}\\
    \phi _{7}^{( e_{1})}( z_{e_1}) & \phi _{7}^{( e_{2})}( z_{e_2}) & \phi _{7}^{( e)}(z_e) =\phi _{4}^{( e_{1})}\\
    \phi _{8}^{( e_{1})}( z_{e_1}) & \phi _{8}^{( e_{2})}( z_{e_2}) & \phi _{8}^{( e)}(z_e) =\phi _{4}^{( e_{2})}\\
    \phi _{9}^{( e_{1})}( z_{e_1}) & \phi _{9}^{( e_{2})}( z_{e_2}) & \phi _{9}^{( e)}(z_e) =\phi _{5}^{( e_{1})}\\
    \vdots  & \vdots  & \vdots 
    \end{pmatrix},
\end{equation}
\noindent and using the symmetrization operator ($\hat{\mathcal{S}}$), we can express the initial wave functions of the $\text{NeHe}^{+}$ ion as an expansion in terms of the previously defined electronic basis as follows:
\begin{equation}
    \Psi _{\text{symm}}^{\text{init}} =\sum _{j=1}^{n} A_{jj1}^{\text{relax}}\hat{\mathcal{S}}\left\{\phi _{( 2j-1)}^{( e)}( z_{e_1}) \phi _{2}^{( e)}( z_{e_2})\right\} \phi _{j}^{( N)},
\end{equation}
\noindent The symmetrized state is then expressed according to the general expansion in Eq.~\eqref{eq:hartree_expansion},
\begin{equation}
    \Psi _{\text{symm}}^{\text{init}} =\sum _{j_{1} ,j_{2} ,j_{3}} A_{j_{1} j_{2} j_{3}}^{\text{symm}} \phi _{j_{1}}^{( e_{1})}( z_{e_1}) \phi _{j_{2}}^{(N)}( R) \phi _{j_{3}}^{( e_{2})}( z_{e_2})
\end{equation}
\noindent where the coefficients $A_{j_{1} j_{2} j_{3}}^{\text{symm}}$ must satisfy the condition that: if $ j_{1} =( 2j_{2} -1)$, $j_{2} =j$ with $j \in \mathbb{N}$ and $ j_{3} =2$ then:
\begin{equation}
    A_{( 2j-1) j 2}^{\text{symm}} =A_{2j( 2j-1)}^{\text{symm}}=A_{jj1}^{\text{relax}}
\end{equation}
\noindent Once the coefficient $j$ is fixed, the value of $A_{j_{1} j_{2} j_{3}}^{\text{symm}}$ is likewise determined.

\section{Propagation setups for MCTDH-Heidelberg package}\label{ap:propagation}

The full three-dimensional system is propagated with 
 the MCTDH algorithm as implemented in Ref.~\cite{mctdh:MLpackage}. There are three DOFs $z_{e_1}$, $R$ and $z_{e_2}$, and we treating the electrons as identical particles (using the same SPFs) in a symmetrical electronic state. The initial state is defined according to the results from the relaxation stage (see Appendix~\ref{ap:initial_state}). The simulations were performed using the parameters listed in Table~\ref{tab:MCTDHprop}.

\begin{table}[h]
    \caption{\label{tab:MCTDHprop} Setting for the propagation of electron-ion scattering using MCTDH-Heidelberg package.}
    \begin{ruledtabular}
        \begin{tabular}{llr}
            Symbol & Description & Value [au]\\
            \hline
            $N_{z_{e}}$ & number of points for DVR grid& 1501 \\
            $N_{R}$ & number of points for DVR grid & 501 \\
            $(z_{e}^{\text{min}},z_{e}^{\text{max}})$ & electronic grid domain range & (-300, 300) \\
            $(R^{\text{min}},R^{\text{max}})$ & nuclear grid domain range & (0, 20) \\
            $\text{SPF}_{z_e}$ & number of electronic SPFs & 20 \\
            $\text{SPF}_{R}$ & number of nuclear SPFs & 18 \\
            $\text{tol}_{\text{CMF/var}}$ & tolerance of CMF integrator & $10^{-7}$ \\
            $\text{tol}_{\text{BS/spf}}$ & tolerance of BS integrator & $10^{-8}$ \\
            $\text{tol}_{\text{SIL/A}}$ & tolerance of SIL integrator & $10^{-7}$ \\
            $t_{\text{final}}$ & final relaxation time & $1446.948$~\footnote{It is equivalent to $35.0$ fs.}\\
            $t_{\text{out}}$ & step relaxation time & $0.413$~\footnote{It is equivalent to $0.01$ fs.}\\
        \end{tabular}
    \end{ruledtabular}
\end{table}

\section{Natural orbital population}\label{ap:nop}

\begin{figure*}
    \centering
    \includegraphics[width=1.0\textwidth]{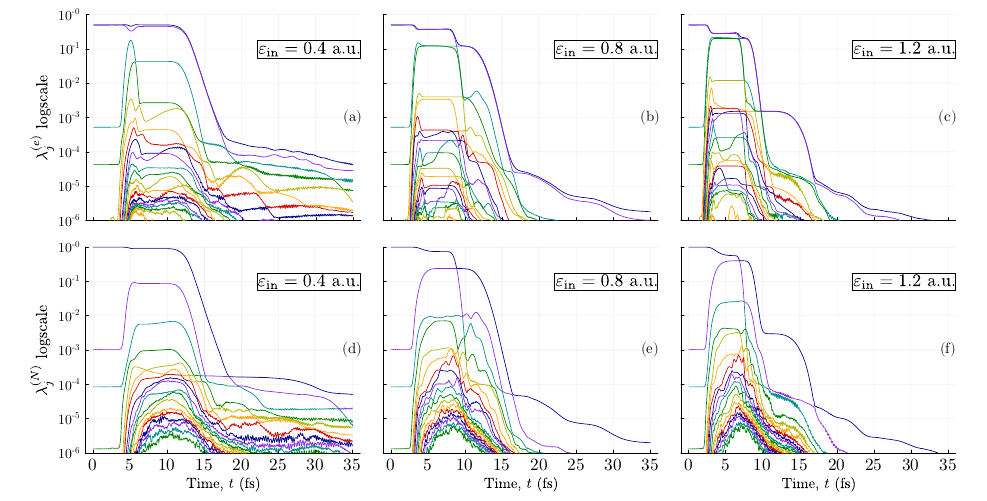}
    \caption{Time evolution of natural orbital populations (NOPs). The top panels show the electronic NOPs ($\lambda^{(e)}_{j}$) and the bottom panels show the nuclear {NOPs} ($\lambda^{(N)}_{j}$). The NOPs line colors highlight the NOPs from the highest one (blue) to the lowest one (yellow). The panels (a) and (d) correspond to an incoming electron energy of $0.4$ a.u., the panels (b) and (e) correspond to an incoming electron energy of $0.8$ a.u. and the panels (c) and (f) correspond to an incoming electron energy of $1.2$ a.u..}
    \label{fig:symm_nat_orb_pop_E4812}
\end{figure*}

Initially, only a few natural orbitals have significant populations, indicating that the initial state of the system is well-described, as we can see in Fig.~\ref{fig:symm_nat_orb_pop_E4812}. We also observe that, for the electronic state, two orbitals are more populated than the others, while for the nuclear state, only one orbital has a higher population than the rest. At the moment of collision, the most populated electronic and nuclear orbitals slightly decrease in population, whereas the less populated orbitals significantly increase in population. This indicates that, at the moment of collision, representing the system’s state becomes more challenging, requiring more coefficients for an accurate description, as seen in Eq.~\eqref{eq:nop}.

After the collision, the natural orbitals that increased in population during the collision begin to decrease, making it easier to represent the state. However, as energy increases, this decrease in orbital population becomes less pronounced. This is because, at higher energies, more ICEC channels are involved, and the inelastic scattering process generates significant correlations in the system, requiring more coefficients to represent the state than in the low-energy case, where only elastic scattering occurs. Finally, at long times, the populations decrease significantly due to the presence of complex absorbing potentials (CAPs), which absorb electronic and nuclear density, causing the orbital norm not to be conserved over time.

    \nocite{*}
    \bibliography{references.bib}              
\end{document}